\documentstyle[12pt]{article} 
\title{\baselineskip=9mm
Validity of the linear coupling approximation in heavy-ion 
fusion reactions at sub barrier energies \\}
\author{K. Hagino$^{*}$ and N. Takigawa 
\\ \\
\medskip
{\it Department of Physics, 
Tohoku University, Sendai 980--77, Japan}
\\ \\
 M. Dasgupta, D.J. Hinde and J.R. Leigh
\\ \\
\medskip
{\it Department of Nuclear Physics,} \\ 
{\it Research School of Physical 
Sciences and Engineering, } \\
{\it Australian National University, Canberra, 
ACT 0200, Australia}
} 
\date{} 

\begin{document}

\maketitle

\vspace*{-1cm}

\begin{center}
{\bf Abstract}
\end{center}

The role of higher order coupling of surface vibrations to 
the relative motion in heavy-ion fusion reactions 
at near-barrier energies is investigated. The coupled channels 
equations are solved to all orders, and also in the
linear and the quadratic coupling approximations. Taking 
$^{64}$Ni + $^{92,96}$Zr reactions as examples, 
it is shown that all order couplings lead to considerably improved 
agreement with the experimentally measured 
 fusion cross sections and  
average angular momenta of the compound nucleus for such heavy nearly 
symmetric systems. 
The importance of higher order coupling is also examined for  
asymmetric systems like $^{16}$O + $^{112}$Cd, $^{144}$Sm, for which 
previous calculations of the fusion cross section seemed to 
indicate that the linear coupling approximation was adequate. It is shown 
that the shape of the barrier distributions and the energy dependence of the 
average angular momentum can change significantly when the 
higher order couplings are included, even for systems where measured 
fusion cross sections may seem to be well reproduced by the linear 
coupling approximation. 

\medskip

\noindent
PACS number(s): 25.70.Jj, 24.10.Eq, 21.60.Ev, 27.60.+j

\medskip

\setlength{\unitlength}{1mm}
\begin{picture}(150,1)
\put(0,0){\line(1,0){150}}
\end{picture}

\noindent
$^{*}$ e-mail address: hagino@nucl.phys.tohoku.ac.jp

\newpage

\begin{center} 
{\bf I. INTRODUCTION}
\end{center}

The analysis of the fusion process in terms of the barrier distribution 
has generated renewed interest 
in heavy-ion collisions at energies below and near the Coulomb barrier
[1--11]. 
Early studies of sub-barrier fusion reactions compared data and theory 
in terms of the excitation function of the fusion cross section. 
It is now well established that fusion cross sections at sub-barrier 
energies may be enhanced by several orders of magnitude compared with 
predictions of a one-dimensional potential model, which is   
due to coupling of the relative motion to nuclear 
intrinsic degrees of freedom\cite{B88}. 
It has been shown that under the eigenchannel approximation, 
these couplings give rise to a distribution of 
potential barriers. 
Recently, Rowley, Satchler, and Stelson proposed a method to extract 
the barrier distribution directly from measured fusion cross 
sections\cite{RSS91}. 
Although, strictly speaking, this method has a clear physical 
meaning only in the limit of sudden tunneling, {\it i.e.} in the limit of 
 a degenerate spectrum of intrinsic motions\cite{HTBB95}, 
the same analysis was later applied also to the case  where the 
intrinsic motions have finite excitation energies\cite{KRNR93}. 
The excitation function of the fusion cross 
section has to be measured with extremely high precision at 
small energy intervals in order to deduce meaningful barrier 
distributions. Such data are now available for several systems, and 
have shown that the barrier distribution is 
very sensitive to the nuclear structure of the colliding 
nuclei. The analysis of the barrier distribution clearly shows 
the effects of couplings to static 
deformations\cite{WLH91,LLW93,LRL93,LDH95}, vibrational 
degrees of freedom\cite{MDH94,LDH95}, transfer channels\cite{MDH94,LDH95}, 
and multi-phonon states\cite{SACN95,SACH95}, as well as the effects 
of projectile excitations\cite{BCL96}, 
in a way much more apparent than in the fusion excitation function itself. 

Another quantity which has recently received increasing attention 
in the study of heavy-ion sub-barrier fusion reactions is the 
angular momentum distribution of the compound nucleus 
[15--23]. 
As in the case of the fusion cross section, the angular momentum 
distribution is also affected by the coupling between the relative 
motion and intrinsic degrees of freedom. 
Experimental data show that the average 
angular momentum of the compound nucleus formed in heavy-ion 
fusion reactions at sub-barrier energies is systematically larger 
than the value expected using the one-dimensional potential model\cite{V92}. 
Furthermore it has been pointed out that the angular momentum distribution 
is also sensitive to the details of the coupling\cite{BBK94}. 

The fusion barrier distribution, the angular momentum distributions 
and the fusion excitation function all reflect the physical processes 
occurring in fusion. In comparing measurement and calculation, the 
excitation function shows most sensitively the energy of the lowest barrier,
whilst the others show more clearly the effect of couplings over the whole 
range of the barrier distribution. Therefore a simultaneous presentation of 
the data and theory in all three forms (if possible) allows the most 
complete comparison of data and theory. 

Theoretically the standard way to address the effects of the 
coupling between 
the relative motion and the intrinsic degrees of freedom is to 
solve the coupled-channels equations, including all the relevant channels. 
Most of the coupled channels calculations performed so far use 
the linear coupling approximation, where the 
coupling potential is expanded in powers of the deformation 
parameter, keeping only the linear term. 
Whilst this approach reproduces the experimental 
data of fusion cross sections for very asymmetric systems, 
it does not explain the data for heavier and nearly symmetric systems
\cite{V92,CBD95,SCA92,ACN94,DNA92}. 
Thus, it is of interest to examine 
the validity of one of the main approximations in these calculations, namely 
the linear coupling approximation, and see whether the effects 
of non-linear coupling improve the agreement between data and 
the theoretical calculations for such systems. 
Even in asymmetric systems, the non-linear couplings might be 
important to reproduce precisely measured data. 

The effects of non-linear coupling can be easily studied 
if the excitation energy of the intrinsic motion is very small 
so as to allow one to use the sudden tunneling approximation\cite{HTBB95}. 
The experimental data of the excitation function of the fusion cross 
section as well as the barrier distribution for the 
$^{16}$O + $^{154}$Sm, $^{186}$W 
reactions were analyzed in this manner\cite{LRL93,LDH95}.  The effects of 
higher order couplings on barrier distributions in the limit of zero 
excitation energy has been discussed by
Balantekin, Bennett, and Kuyucak in the framework of the 
interacting boson model(IBM) \cite{BBK93}. 
However, for nuclear surface vibrations the excitation energies 
cannot be neglected in most cases, and one has to solve 
full coupled channels equations. Due to the complexity  of such 
calculations,  very  few studies have addressed the effects of the higher 
order couplings for the vibrational motion. Esbensen and Landowne expanded 
the coupling potential up to the second order with respect to 
the deformation parameter, and have shown that second order coupling 
leads to a better agreement between the theoretical calculations 
and the experimental fusion 
cross sections for reactions between different nickel 
isotopes\cite{EL87}. The quadratic coupling approximation 
was applied also to the $^{58,64}$Ni + $^{92,100}$Mo reactions\cite{REG93}. 
There it was shown that the experimental data of both 
the fusion cross sections and the angular momentum distributions are 
well reproduced 
by the coupled channels calculations in the quadratic coupling 
approximation. 
Coupled channels calculations including full order 
 coupling and the finite excitation energy of nuclear surface 
vibrations have recently been performed  
for the $^{58}$Ni + $^{60}$Ni reaction\cite{SACN95,R95}. It is seen  
that higher order 
couplings are essential in reproducing the experimental data for this 
system, and the shape of the barrier distribution changes drastically 
 when the effects of the higher order couplings are taken into 
account. 
Refs.\cite{SACN95,R95} do not, however, discuss 
the quality of the quadratic coupling approximation and the convergence 
of the expansion of the coupling potential.

Although all the above studies and the multi-dimensional tunneling
 model in Ref.\cite{DR95} show 
the effects of higher order couplings in specific systems,  
there has not been  any systematic effort to identify their degree of 
importance for different systems. Furthermore, 
it is not obvious whether calculations to 
all orders are necessary or the  expansion up to the second order is 
sufficient. In view of the high precision data that have recently become 
available, a critical examination of the effects of these approximations 
on the cross-sections and barrier distribution is necessary before making
quantitative comparisons with experimental data.

In this paper we solve the coupled channels equations including the 
finite excitation energies of the vibrational states, and without 
introducing the expansion  with respect to deformation parameters. 
The results of these calculations for fusion cross sections, average angular 
momenta and barrier distributions are compared with
those using the linear and the quadratic coupling approximations.
The paper is organized as follows. 
In Sec. II  the coupled channels calculations 
which include higher order couplings are formulated. 
Explicit expressions for the matrix elements of higher 
order terms in both the nuclear and the Coulomb couplings are 
presented. It is seen that inclusion of up to the first second order term in 
 the Coulomb  coupling is sufficient, but  higher order terms 
are necessary for nuclear coupling. 
In Sec. III the coupled channels equations are solved for the 
$^{64}$Ni + $^{92,96}$Zr systems, 
where the experimental data of both the fusion cross sections and the 
average angular momenta of the compound nucleus are available. 
The asymmetric systems $^{16}$O + $^{112}$Cd, $^{144}$Sm 
are also studied in this section and the calculations are compared with
measured fusion cross sections, barrier distributions, 
and the average angular momenta of the compound nucleus. 
The summary is given in Sec. IV. 

\begin{center}
{\bf II. COUPLED CHANNELS EQUATIONS}\\
{\bf AND COUPLING FORM FACTORS}
\end{center}

Consider the problem where the relative motion between  
colliding nuclei couples to a vibrational mode of excitation of the target 
nucleus. For simplicity excitations of the 
projectile are not considered in this section. It is straightforward 
to extend the formulae to the case 
where many different vibrational modes are present and where projectile 
excitations also occur. For heavy ions, to a good approximation 
one can replace the angular momentum of the relative motion in
each channel by the total angular momentum $J$\cite{TI86,HTBBe95}. This 
approximation, often referred to as 
no-Coriolis approximation,  will be used throughout this 
paper. The coupled channels equations then read
\begin{equation}
\left[-\frac{\hbar^2}{2\mu}\frac{d^2}{dr^2}
+\frac{J(J+1)\hbar^2}{2\mu r^2}
+V_N(r)+\frac{Z_PZ_Te^2}{r}+n\hbar\omega
-E_{cm}\right]\psi_n(r)+\sum_mV_{nm}(r)\psi_m(r)=0  ,
\end{equation}
where $r$ is the radial component of the coordinate of the 
relative motion, $\mu$ the reduced mass, and $V_N$ the nuclear potential 
in the entrance channel, respectively. $E_{cm}$ is the
bombarding energy in the center of mass frame and 
$\hbar\omega$ is the excitation energy of 
the vibrational phonon.
$V_{nm}$ are the coupling form factors, which in the collective model 
consist of Coulomb and nuclear components. These two components are 
discussed in the following sub-sections.

\begin{center}
{\bf A. Coulomb coupling form factors}
\end{center}

We first consider the effects of 
higher order terms of the Coulomb component. In Refs.\cite{BBK93,EL87} 
it has been reported that the higher order Coulomb 
couplings are not important in heavy ion fusion reactions. 
However, Ref.\cite{EL87} studied only the excitation function of the 
fusion cross section, and did not discuss the barrier distribution. 
On the other hand, Ref.\cite{BBK93} ignored the finite excitation energy 
of nuclear intrinsic motions, though it discusses 
the effects on barrier distribution. 
Here we investigate the effects of higher order 
Coulomb couplings on both the excitation function of the fusion 
cross section and the barrier distribution and do not ignore the energy 
of nuclear intrinsic excitations. 
Initially, we consider the case where 
the target has only a single phonon 
excitation. 

The Coulomb potential 
between the spherical projectile and the vibrational target is given by 
\begin{equation}
V_C(r)=\left.\int d{\mbox{\bf r}}' 
\frac{Z_PZ_Te^2}{|{\mbox{\bf r}}-{\mbox{\bf r}}'|}
\rho_T({\mbox{\bf r}}') \right|_{\hat{{\mbox{\bf r}}}=0}
=\frac{Z_PZ_Te^2}{r}+\sum_{\lambda'\ne 0}\frac{4\pi Z_Pe}{2\lambda'+1}
\sqrt{\frac{2\lambda'+1}{4\pi}}Q_{\lambda'0}\frac{1}{r^{\lambda'+1}} ~~~~,
\end{equation}
where $\rho_T$ is the charge density of the target nucleus 
and $Q_{\lambda'0}$ the electric multipole operator defined as 
\begin{equation}
Q_{\lambda'0}=\int d{\mbox{\bf r}} Z_Te\rho_T({\mbox{\bf r}})
r^{\lambda'}Y_{\lambda'0}(\hat{{\mbox{\bf r}}})  .
\end{equation}
Eq.(2) uses the fact that the angular momentum for the relative 
motion does not change in the no-Coriolis approximation, and that 
the associated spherical harmonics are evaluated at the forward angle 
${\hat{{\mbox{\bf r}}}=0}$, leading to the factor 
$\sqrt{\frac{2\lambda'+1}{4\pi}}$. 
If we assume a sharp matter distribution for the target nucleus and 
a phonon excitation of multipolarity $\lambda$, then 
the electric multipole operator is given by\cite{RS80}
\begin{equation}
Q_{\lambda'0}=\frac{3e}{4\pi}Z_T\left(R_T^{(0)}\right)^{\lambda'}
\left\{\alpha_{\lambda 0}\delta_{\lambda,\lambda'}
+(-)^{\lambda'}\frac{(\lambda'+2)(2\lambda+1)}{2\sqrt{4\pi}}
\left(\begin{array}{ccc}
\lambda&\lambda&\lambda'\\
0&0&0
\end{array}
\right)
(\alpha_{\lambda}\alpha_{\lambda})_{\lambda'0}\right\} ~~~,
\end{equation}
up to second order in the surface coordinate 
$\alpha_{\lambda\mu}$, where $R_T^{(0)}$ 
is the equivalent sharp surface radius of the 
target nucleus. In the collective model of surface oscillations, the surface 
coordinates $\alpha_{\lambda\mu}$ are treated as dynamical variables. 
They are related to the phonon creation and annihilation operators by 
\begin{equation}
\alpha_{\lambda\mu}=\alpha_0(a^{\dagger}_{\lambda\mu}
+(-)^{\mu}a_{\lambda-\mu}),
\end{equation}
where $\alpha_0$ is the amplitude of the zero point motion. 
It is related to the deformation parameter $\beta_{\lambda}$ 
by $\alpha_0=\beta_{\lambda}/\sqrt{2\lambda+1}$\cite{BM75} 
and can be estimated from the measured transition probability using 
\begin{equation}
\alpha_0=\frac{1}{\sqrt{2\lambda+1}}\frac{4\pi}{3Z_T(R_T^{(0)})^{\lambda}}
\sqrt{\frac{B(E\lambda)\uparrow}{e^2}}~~~~~~~.
\end{equation}
This equation is valid if the amplitude of the vibration is small, 
and the transition operator is linearly proportional to $\alpha_0$. 

The Coulomb components of the coupling form factors $V_{nm}$ in 
Eq.(1) are obtained by taking the matrix elements of $V_C$ 
between $n$- and $m$- phonon states. 
Since we assume that there exists only the one phonon state in the 
vibrational excitation of the target, 
the Coulomb coupling form factors up to second order 
of $\alpha_0$ are given by 
\begin{eqnarray}
V^{(C)}_{01}(r)&=&V^{(C)}_{10}(r)=\frac{3}{2\lambda+1}Z_PZ_Te^2
\frac{(R_T^{(0)})^{\lambda}}{r^{\lambda+1}}
\sqrt{\frac{2\lambda+1}{4\pi}}\alpha_0 \\
V^{(C)}_{11}(r)&=&2\sum_{\lambda'\ne0}(-)^{\lambda'}
\frac{3(2\lambda+1)(\lambda'+2)}{8\pi(2\lambda'+1)}
<\lambda 0 \lambda 0|\lambda'0>^2\alpha_0^2Z_PZ_Te^2
\frac{(R_T^{(0)})^{\lambda'}}{r^{\lambda'+1}}.
\end{eqnarray}
If there exist two-phonon multiplets, then the formalism becomes much 
more complicated in the case of non-linear coupling. 
In the case of the linear coupling, it is known that 
the no-Coriolis approximation enables 
us to replace the couplings to all the  members of the two-phonon 
multiplets by the coupling to a single state 
by making an appropriate unitary transformation 
\cite{KRNR93,TI86}. 
This leads to a significant reduction of the dimensions of the 
coupled channels problem. This property is lost 
if one keeps higher order terms of the Coulomb coupling since 
 the radial dependence of 
the coupling form factor for the Coulomb part 
explicitly depends on the multipolarity of the nuclear excitation. 

We now apply Eqs. (7) and (8) to fusion reactions between 
two $^{58}$Ni nuclei, where the importance of second order 
couplings in the nuclear interaction has been reported\cite{EL87}. 
We take into account the quadrupole vibrational state at 1.45 MeV, 
and truncate the whole space at the one phonon state. 
The parameters for the nuclear potential and 
the deformation 
parameter from Ref.\cite{EL87} have been used. 
Since at this stage 
we want to investigate the effects of higher order Coulomb coupling, a 
linear coupling for the nuclear interaction has been used, for  ease of 
calculation. 
The coupled channels equations are solved by imposing the 
incoming wave boundary condition in the inner region 
of the fusion potential. 
We found that the second order coupling in the 
Coulomb interaction causes no visible effects to the fusion cross 
section. It changes the fusion cross section by only about 0.2 \% 
in the energy region we considered, {\it i.e.} from about 10 MeV below 
the Coulomb barrier to about 10 MeV above the Coulomb barrier. 
Fig.1 shows the barrier distribution 
($d^2(E_{c.m.}\sigma)/dE_{c.m.}^2$ ) as a function of the bombarding energy. 
As seen from the figure, second order Coulomb couplings 
modify only very marginally the barrier distribution. 
Further calculations showed that the situation does not change when 
the value of the deformation parameter 
is varied within physically plausible limits. 
Therefore we hereafter use the linear coupling approximation for the 
Coulomb coupling and investigate the effects of the higher order 
terms only for nuclear coupling. 
The matrix elements of the Coulomb coupling form factor 
in Eq. (1) are now given by 
\begin{eqnarray}
V^{(C)}_{nm}(r)&=&\frac{3}{2\lambda+1}Z_PZ_Te^2
\frac{(R_T^{(0)})^{\lambda}}{r^{\lambda+1}}
\sqrt{\frac{2\lambda+1}{4\pi}}\alpha_0 (\sqrt{n}\delta_{n,m+1}
+\sqrt{n+1}\delta_{n,m-1})\\
&=&\frac{3}{2\lambda+1}Z_PZ_Te^2
\frac{(R_T^{(0)})^{\lambda}}{r^{\lambda+1}}
\frac{\beta_{\lambda}}{\sqrt{4\pi}} (\sqrt{n}\delta_{n,m+1}
+\sqrt{n+1}\delta_{n,m-1}).
\end{eqnarray}
Note that we have defined the multi-phonon channels by taking the 
appropriate linear combinations of the multi-phonon multiplets. 
As remarked before, this is possible only for the 
linear coupling approximation in the Coulomb interaction. 

\begin{center}
{\bf B. Nuclear coupling form factors}
\end{center}

In the collective model, the nuclear interaction is assumed to be 
a function of the separation distance between the vibrating surfaces 
of the colliding nuclei. 
It is conventionally taken as 
\begin{equation}
V^{(N)}(r,\alpha_{\lambda0})
=-\frac{V_0}{1+\exp[(r-R_P-R_T^{(0)}-\sqrt{\frac{2\lambda+1}{4\pi}}
R_T^{(0)}\alpha_{\lambda0})/a]}.
\end{equation}
Volume conservation introduces a small term which is non-linear with respect 
to the deformation parameter $\alpha_{\lambda0}$ in the denominator of the 
above Eq.(11). This is ignored for simplicity in the present study.
As in the case of Eq.(2) for the Coulomb coupling, here we consider 
the coupling form factor for the forward angle, which is needed 
to obtain the coupled channels equations in the no-Coriolis approximation. 
We assume a Woods-Saxon 
form for the nuclear potential. The structure of the 
resultant formulae in this subsection, 
however, remain unchanged for other forms of the nuclear potential. 
Denoting the eigenvalue of $\alpha_{\lambda0}$ by $x$, 
the matrix elements of the nuclear coupling form factor read 
\begin{equation}
V_{nm}^{(N)}(r)
=\int^{\infty}_{-\infty}dxu^*_n(x)u_m(x)
\frac{-V_0}{1+\exp[(r-R_P-R_T^{(0)}-\sqrt{\frac{2\lambda+1}{4\pi}}
R_T^{(0)}x)/a]}~~.
\end{equation}
$u_n(x)$ is the eigen function of the $n$ -th excited state of the 
harmonic oscillator. 
The conventional nuclear coupling form factor in the linear coupling 
approximation is obtained by expanding Eq.(11) with respect to 
$\alpha_{\lambda0}$ and keeping only the linear term. 

The expectation value of the nuclear potential in the ground 
state is often replaced by the phenomenological potential 
\begin{equation}
V_N(r)=-\frac{V_0}{1+\exp[(r-R_P-R_T^{(0)})/a]}~~,
\end{equation}
which is assumed to be known empirically\cite{EL87}. 
If we take this prescription, the nuclear coupling form factor 
in Eq.(1) is calculated as 
\begin{eqnarray}
V_{nm}^{(N)}(r)
&=&\int^{\infty}_{-\infty}dxu^*_n(x)u_m(x)
\frac{-V_0}{1+\exp[(r-R_P-R_T^{(0)}-\sqrt{\frac{2\lambda+1}{4\pi}}
R_T^{(0)}x)/a]} \nonumber \\
&&-\delta_{n,m}\int^{\infty}_{-\infty}dx|u^*_0(x)|^2
\frac{-V_0}{1+\exp[(r-R_P-R_T^{(0)}-\sqrt{\frac{2\lambda+1}{4\pi}}
R_T^{(0)}x)/a]} ~~~~.
\end{eqnarray}
The last term in this equation is included to make the coupling 
interaction vanish in the entrance channel. 
Eq. (14) represents the form factor which 
contains couplings to all orders.  
We use these form factors in the next section 
in order to discuss the effects of higher order coupling to vibrational 
modes of excitation of the colliding nuclei on heavy ion fusion 
reactions. 


Instead of introducing a phenomenological potential given by Eq.(13) as 
the bare potential in the entrance channel, one could use 
$V_{00}^{(N)}$ in Eq. (12) as the nuclear potential in the entrance 
channel. The use of Eq.(13) makes it easier to 
examine the convergence of the effects of 
higher order terms 
by comparing the results of the 
calculations in the linear and the  
quadratic approximations and the full order calculations. 
Notice that $V_{00}^{(N)}$ is identical with the potential 
given by Eq.(13) in the linear coupling approximation. 

\begin{center}
{\bf III. RESULTS: EFFECTS OF HIGHER ORDER COUPLINGS}
\end{center}

\begin{center}
{\bf A. Nearly symmetric systems}
\end{center}

We now present the results of our calculations of fusion cross sections, 
average angular momenta of the compound nucleus, and fusion barrier 
distributions. 
We first discuss heavy nearly symmetric systems. 
The experimental data of the average angular momentum of the compound nucleus 
for several systems are summarized in Fig.5 of Ref. \cite{CBD95}.  
It suggests that the conventional coupled channels 
calculations do not work for heavy 
nearly symmetric systems. 
We analyse in particular $^{64}$Ni + $^{92,96}$Zr 
reactions which are typical examples where 
the conventional coupled channels calculations with the linear 
coupling approximation fail to reproduce the 
fusion cross sections and average angular momentum data\cite{SCA92}. 
Our aim is to investigate whether 
the failure is due to the linear coupling approximation by 
performing linear, quadratic and full coupling calculations. 

We take into account the couplings up to two phonon states 
of the quadrupole surface vibrations of 
$^{64}$Ni and $^{92}$Zr, and of the octupole vibration of 
$^{96}$Zr. We also take their mutual excitations into account. 
The excitation energies of the single phonon states in $^{64}$Ni, $^{92}$Zr, 
and $^{96}$Zr are 1.34, 0.934, and 1.897 MeV, respectively. 
We assumed the radius parameter 
associated with the coupling interactions to be 1.2 fm in all cases. 
The deformation parameter of $^{64}$Ni was taken 
to be $\beta_2=0.19$ \cite{EL87}. 
Following Refs.\cite{KRNR93,TSTKD90} we used $\beta_2=0.25$ for the nuclear 
coupling associated with the quadrupole vibration of $^{92}$Zr, while 
the deformation parameter in the Coulomb coupling interaction was estimated 
from the measured B(E2)$\uparrow$ value to be 0.108. 
The different value for the nuclear deformation parameter 
from that of the Coulomb coupling parameter was required in order 
 to fit the angular distribution of the inelastic scattering 
of $^{16}$O from $^{92}$Zr at 56 MeV\cite{TSTKD90}. 
The deformation parameter $\beta_3$ of $^{96}$Zr was estimated from 
the recently measured B(E3)$\uparrow$ value\cite{HAS93} to be 0.268. 
We assumed the same value for the deformation parameter 
for the nuclear coupling as 
that for the Coulomb deformation parameter for this nucleus. 
The nuclear potentials used in this paper are the same as in 
Ref.\cite{SCA92}. 
These modify the empirical potentials of Christensen and Winther\cite{CW76} 
by setting the range adjustment parameter $\Delta R$ to be 0 fm. 

The excitation function of the fusion cross section for these two systems 
obtained by numerically solving 
the coupled channels equations are compared with the experimental data 
in Figs. 2 and 3 (upper panels). The experimental data, 
taken from Ref. \cite{SCA92}, consist only 
of the evaporation residue cross sections, 
and do not include fission following fusion. 
The dotted lines are the results in the one dimensional 
potential model, {\it i.e. } without the effects of channel coupling. 
As is well known, the experimental fusion 
cross sections at subbarrier energies are several orders of 
magnitude larger than the predictions of this model. 
The dot-dashed lines are the results of the coupled 
channels calculations when the linear coupling approximation is used, 
which are similar to the results of the simplified coupled channels 
calculations reported in Ref.\cite{SCA92}. 
They considerably underestimate the fusion cross sections 
at sub-barrier energies for both systems. 
The situation is slightly improved when the 
quadratic coupling approximation is used, {\it i.e.} when 
the nuclear coupling potential up to the second order of the deformation 
parameter\cite{EL87}, is included (dashed lines).  
However, there still remain considerable discrepancies between the 
experimental data and the results of the coupled channels calculations. 
When we include  couplings to all order, we get the solid lines, which 
agree very well with the experimental data. Dramatic effects 
of the higher order couplings on fusion cross sections are 
observed, especially at low energies. 

The lower panels in Figs. 2 and 3 compare the 
results of our calculations of the average angular momentum 
of the compound nucleus with the experimental data 
as a function of the bombarding energy.
It is defined in terms of the partial fusion cross section $\sigma_l$ as 
\begin{equation}
<l>=\sum_ll\sigma_l/\sum_l\sigma_l
\end{equation}
The meaning of each line in these figures 
is the same as in the upper panels. 
We again observe that the experimental data are much better reproduced 
by taking the effects of couplings to all orders into account. 
We thus conclude that coupling to all 
orders are essential to simultaneously reproduce the fusion cross 
sections and the average angular momentum data 
for heavy (nearly) symmetric systems. This is in agreement with the 
calculations required to fit the barrier distribution for 
$^{58}$Ni + $^{60}$Ni reaction\cite{SACN95}. 

\begin{center}
{\bf B. Very asymmetric systems}
\end{center}

We next consider the effects of higher order couplings for 
very asymmetric systems where the product of the charges $Z_PZ_T$ 
 is relatively small. 
For such systems, the coupled channels calculations in the linear 
coupling approximation have achieved reasonable success 
in reproducing fusion excitation  functions.
However, no study has been performed to see whether 
the effects of higher order couplings on the angular 
momentum distribution of the compound nucleus and on the barrier 
distributions are small. In this subsection we re-analyse 
the experimental data for the $^{16}$O + $^{112}$Cd reaction, for which 
both fusion cross sections and average angular momentum data are 
available\cite{ACN94}, 
and those for the $^{16}$O + $^{144}$Sm reaction, where the 
the fusion barrier distribution has been extracted from 
the precisely measured fusion cross sections\cite{LDH95}. 
For simplicity in the calculations we ignore  excitation of the projectile 
in both reactions. These effects will be discussed in a separate 
paper\cite{HTLD96}, where it will be shown that the 
octupole vibration of $^{16}$O leads to a static 
renormalization of the fusion barrier\cite{THAB94,THA95}. 

In calculating the fusion cross section for  $^{16}$O + $^{112}$Cd 
scattering, 
we include the double quadrupole phonon states and the single octupole phonon 
state of $^{112}$Cd and their mutual excitations. 
The excitation energies are 0.617 and 2.005 MeV for 
the one phonon states of the quadrupole and the octupole 
vibrations, respectively. 
The deformation parameters of the quadrupole and the 
octupole vibrations are estimated to be $\beta_2=0.173$ and $\beta_3=0.164$, 
respectively\cite{ACN94}. 
The radius parameter in the 
coupling interaction is taken to be 1.2 fm. 
Following Ref.\cite{ACN94}, we use a Woods-Saxon potential 
whose depth, range parameter, and  surface diffuseness are 
$V$=58 MeV, $r_0$=1.22 fm, and $a$=0.63 fm, respectively. 

The upper panel of Fig. 4 compares the 
results of the coupled channels 
calculations of fusion cross sections with the 
experimental data taken from Ref.\cite{ACN94}. 
Compared with the symmetric systems studied in the previous 
subsection, the enhancement of the fusion cross sections is 
fairly small. 
This is partly because the product of the atomic number 
$Z_PZ_T$ in this asymmetric system is smaller than the symmetric systems. 
If we take the linear coupling approximation and estimate 
the coupling strength $F$ at the barrier position $r_B$ 
of the bare Coulomb barrier, one finds
\begin{equation}
F=\frac{\beta_{\lambda}}{\sqrt{4\pi}}Z_PZ_Te^2\left(
-\frac{R_T^{(0)}}{r_B^2}+\frac{3}{2\lambda+1}
\frac{(R_T^{(0)})^{\lambda}}{r_B^{\lambda+1}}\right).
\end{equation}
The coupling strength is thus proportional to the 
product $Z_PZ_T$. 
This product is 384 for $^{16}$O + $^{112}$Cd scattering, whilst it is 
1120 for $^{64}$Ni + $^{92,96}$Zr reactions. The coupling strength 
in this asymmetric system is therefore several times smaller 
than in the symmetric systems even though the values of the 
deformation parameters are similar. 
Another reason that 
the enhancement of the fusion cross sections is small in very 
asymmetric systems is the small reduced mass.
In the WKB formula for the barrier penetrability, 
the mass parameter appears in the exponent. 
Hence the heavier the mass, the more sensitive the penetrability 
to a slight change of the potential. 
Even though the results in the linear coupling approximation (dot-dashed
line) show a relatively small enhancement 
of the fusion cross section
compared with the no-coupling
limit, there is still a significant change in going to second order
coupling, and then to all order coupling. 
The situation is similar for the average angular momentum. 
Thus even in such cases with low
$Z_PZ_T$, if data of high precision are available, it seems that the linear
coupling approximation is inadequate to allow quantitative conclusions to be
drawn from a comparison of data and calculations.

The role of higher order couplings in very asymmetric systems
can be more clearly seen by investigating the fusion barrier distributions.
Therefore, we next consider the $^{16}$O + $^{144}$Sm reaction,
for which the effects on fusion barrier distributions of couplings to 
phonon states were shown experimentally 
for the first time\cite{MDH94,LDH95,RTL96}.
The authors of Ref.\cite{LDH95}
have shown that the fusion barrier distribution for this system is
intimately related to the octupole vibration of $^{144}$Sm, and that
the quadrupole vibration plays only a minor role.
Accordingly, we ignore the effects of the
couplings to the quadrupole phonon states of $^{144}$Sm and include only the
single octupole phonon state at 1.81 MeV.
The deformation parameter $\beta_3$= 0.205 was used as in  Ref.\cite{LDH95}. 
The ion-ion potential was of a Woods-Saxon form. The depth, 
radius parameter and surface diffuseness were 105.1 MeV, 1.1 fm and 0.75 fm 
respectively, as given in Ref.\cite{M95}. 

The upper panel of Fig. 5 shows the fusion excitation function 
from Ref.\cite{LDH95}, and the calculations.  The meaning of each line 
is the same as in Fig.2. As was the case for $^{16}$O + $^{112}$Cd, 
we observe that the
agreement of the theory and experiment 
appears to be improved only slightly by the inclusion of coupling to all 
orders.
The barrier distribution, however, reveals significant changes due to
the higher order couplings( see the lower panel of Fig.5).
Note that there exist two barriers in the present two channel problem.
Comparing the results of the linear coupling approximation( the dot-dashed
line) with those of the all order coupling( the solid line), one observes
that the higher order couplings 
transfer some strength from the lower barrier 
to the higher barrier,
and at the same time lower the energies of both barriers.

This can be viewed in a different way by performing the diagonalization
of the coupling matrix at each position of the internuclear separation
to obtain the effective barriers, as is done in the computer code
CCMOD\cite{DNA92}.
Fig. 6 shows these effective barriers for s-wave scattering. 
The meaning of each line is the same as in Fig.5.
We observe that higher order couplings decrease the energies of both the lower
and the higher barriers, consistent with the barrier distributions shown 
in Fig. 5. 
The higher order couplings also increase the width of both potential 
barriers (Fig.6), leading to narrower peaks in the barrier distribution. 
This then results in
the apparent better separation between the two barriers seen in Fig. 5. 

%
%

For these asymmetric
reactions, the couplings are weak as a result of a
combination of the small product of $Z_PZ_T$ and the
relatively small deformation parameters.
In such cases the
first order approximation might have been expected to be valid. Despite
this, the calculations which include couplings to all orders 
show significant differences from first order calculations.
It is clear
therefore that high precision measurements should be analysed using all
order couplings even when coupling is weak.

\begin{center}
{\bf IV. SUMMARY}
\end{center}

We have shown that  in heavy ion fusion reactions, higher order couplings
to nuclear surface vibrations play an important role. Such higher 
order terms in the Coulomb coupling can be safely neglected. Previous work 
indicated that standard
coupled channels calculations are not very successful in describing fusion
of heavy symmetric
systems. We have shown that the data can be described well by
coupled channels calculations
 once couplings to all orders are included.
We found that for the $^{64}$Ni + $^{92,96}$Zr reactions, terms beyond those
 in the quadratic coupling approximation result in further enhancement of
the fusion cross sections at subbarrier energies.
The additional enhancement is as large as that due to the inclusion of
quadratic coupling. The inclusion of the coupling to all orders is
crucial to reproduce the experimental
 fusion cross sections and the average angular momenta.
We performed calculations also for the $^{16}$O + $^{112}$Cd, $^{144}$Sm
reactions as examples of very asymmetric systems where the coupling
is weaker. It is found that in such cases higher order couplings result 
in a non-negligible enhancement of the  fusion cross sections and a 
significant modification of barrier distributions as well as the average 
angular momenta.

High precision fusion cross section measurements to deduce the barrier
distribution, and measurements of angular momentum
distributions, are designed to study  the important couplings in a reaction.
The sensitivity to couplings is greatly enhanced  by performing experiments
involving target projectile combinations with a large value of $Z_PZ_T$. 
It has been shown in this paper that higher order
coupling significantly affects the barrier distribution and average angular
momentum even for weak coupling cases like $^{16}$O + $^{112}$Cd, $^{144}$Sm
 with values of $Z_PZ_T \sim$400. 
Thus, spurious conclusions regarding the nature of couplings could 
be reached if high quality experimental data, particularly for heavier 
systems, are compared with calculations performed only with first 
order coupling. 
The  stage has now been reached when the standard codes of the 
coupled channels calculations should be revised to include 
coupling to all orders.

\begin{center}
{\bf ACKNOWLEDGMENTS} 
\end{center}

The authors thank S. Kuyucak, J.R. Bennett and M. Abe for useful discussions.
K.H. and N.T. also thank the Australian National University for its 
hospitality and for partial support for this project.
The work of K.H. was supported by the Japan Society for the Promotion 
of Science for Young Scientists.
This work was supported by the Grant-in-Aid for General
Scientific Research,
Contract No.06640368, and the Grant-in-Aid for Scientific
Research on Priority Areas, Contract No.05243102, 
from the Japanese Ministry of Education, Science and Culture, 
and a bilateral program of JSPS between Japan and Australia. 


\begin{center}
{\bf Figure Captions}
\end{center}

\noindent
{\bf Fig.1:}The barrier distribution for fusion 
between two $^{58}$Ni nuclei.
The one phonon state of the quadrupole surface vibration
is taken into account.
The nuclear interaction is treated in the linear coupling approximation.
The dotted line corresponds to the case where the
Coulomb coupling potential is also treated in the
linear coupling approximation, while the solid line takes into
account non-linear terms up to the second order.

\noindent
{\bf Fig.2:}Excitation function of the fusion cross section (upper
panel) and the average angular momentum of the compound nucleus
(lower panel) for the 
$^{64}$Ni + $^{92}$Zr reaction. 
The experimental data are taken
from Ref.\cite{SCA92}.
The two phonon states of the quadrupole surface vibration
of both the projectile and the target are taken into account
in the coupled channels calculations.
The dotted line is the result in the absence of channel coupling.
The dot-dashed and the dashed lines are the results when the nuclear
potential is expanded up to the first and the second order terms in the
deformation parameters, respectively. The solid line is the results of the
coupled channels calculations to all orders, obtained
without expanding the nuclear potential.

\noindent
{\bf Fig.3:}Same as Fig.2, but for $^{64}$Ni + $^{96}$Zr fusion.
The two phonon states of the quadrupole surface vibrations
of the projectile and those of the octupole
surface vibrations of the target are taken into account
in the coupled channels calculations.
The experimental data are taken from Ref.\cite{SCA92}.

\noindent
{\bf Fig.4:}Same as Fig.2, but for $^{16}$O + $^{112}$Cd fusion.
In the coupled channels calculations,
the projectile is assumed to be inert. The one and two
 phonon quadrupole states and the one phonon octupole state of the target are
taken into account.
The experimental data are taken from Ref.\cite{ACN94}.

\noindent
{\bf Fig.5:}Excitation function of the fusion cross section (upper
panel) and the barrier distribution (lower panel) for
$^{16}$O + $^{144}$Sm fusion. In the coupled channels calculations,
the projectile is assumed to be inert, while the single octupole
phonon state of the target is taken into account.
The meaning of each line is the same as in Fig.2.
The experimental data are taken from Ref.\cite{LDH95}.

\noindent
{\bf Fig.6:}
Effective potential barriers for the s-wave scattering of
$^{16}$O from $^{144}$Sm  obtained by
diagonalizing the coupling matrix.
The meaning of each line is the same as in Fig.5.

\end{document}